\documentclass[letter]{IEEEtran}
\IEEEoverridecommandlockouts
\usepackage{cite}
\usepackage{amsmath,amssymb,amsfonts}
\usepackage{amssymb}
\usepackage{algpseudocode}
\usepackage{amsfonts}
\usepackage{graphicx}
\usepackage{fancyhdr}  %
\usepackage{cases}
\usepackage{textcomp}
\usepackage{extarrows}
\usepackage{algorithm,algpseudocode}
\usepackage{multirow}
\usepackage{booktabs}
\usepackage{caption}
\usepackage{subcaption}
\usepackage{orcidlink}

\usepackage{multirow,tabularx}
\usepackage{mathtools}
\usepackage{xcolor}
\usepackage[english]{babel}
\usepackage{amsthm}
\usepackage[left]{lineno}

\hypersetup{hidelinks}

\usepackage{caption}
\def\BibTeX{{\rm B\kern-.05em{\sc i\kern-.025em b}\kern-.08em
    T\kern-.1667em\lower.7ex\hbox{E}\kern-.125emX}}
\begin{document}
\bstctlcite{BSTcontrol}
\title{Robust Beamforming with Gradient-based Liquid Neural Network}

\author{
	\IEEEauthorblockN{
	Xinquan Wang$^{\orcidlink{0009-0005-9986-7054}}$,
	Fenghao Zhu$^{\orcidlink{0009-0006-5585-7302}}$,
	Chongwen Huang$^{\orcidlink{0000-0001-8398-8437}}$,
	Ahmed Alhammadi$^{\orcidlink{0009-0006-8535-268X}}$,
	Faouzi Bader$^{\orcidlink{0000-0002-3206-9411}}$,\\
	Zhaoyang Zhang$^{\orcidlink{0000-0003-2346-6228}}$,
	Chau Yuen$^{\orcidlink{0000-0002-9307-2120}}$,~\IEEEmembership{Fellow,~IEEE}, and
	M\'{e}rouane~Debbah$^{\orcidlink{0000-0001-8941-8080}}$,~\IEEEmembership{Fellow,~IEEE}}
\thanks{The work was supported by the China National Key R\&D Program under Grant 2021YFA1000500 and 2023YFB2904804, National Natural Science Foundation of China under Grant 62331023, 62101492, 62394292 and U20A20158, Zhejiang Provincial Natural Science Foundation of China under Grant LR22F010002, Zhejiang Provincial Science and Technology Plan Project under Grant 2024C01033, and Zhejiang University Global Partnership Fund, MOE Tier 2 (Award number MOE-T2EP50220-0019) and A*STAR (Agency for Science, Technology and Research) Singapore, under Grant No. M22L1b0110. (\textit{Corresponding author: Chongwen Huang.})}

\thanks{X. Wang and C. Huang are with College of Information Science and Electronic Engineering, Zhejiang University, Hangzhou 310027, China, the State Key Laboratory of Integrated Service Networks, Xidian University, Xi’an 710071, China, and Zhejiang Provincial Key Laboratory of Info. Proc., Commun. \& Netw. (IPCAN), Hangzhou 310027 China (E-mails: \href{mailto:wangxinquan@zju.edu.cn}{\{wangxinquan}, \href{mailto:chongwenhuang@zju.edu.cn}{chongwenhuang\}@zju.edu.cn}).}
\thanks{
F. Zhu and Z. Zhang are with the College of Information Science and Electronic Engineering, Zhejiang University, Hangzhou 310027, China (E-mails: \href{mailto:zjuzfh@zju.edu.cn}{\{zjuzfh}, \href{mailto:yang_zhaohui@zju.edu.cn}{yang\_zhaohui}, \href{mailto:chen_xiaoming@zju.edu.cn}{chen\_xiaoming}, \href{mailto:ning_ming@zju.edu.cn}{ning\_ming\}@zju.edu.cn}).}
\thanks{A. Alhammadi and F. Bader are with Technology Innovation Institute, 9639 Masdar City, Abu Dhabi, UAE (E-mails: \href{mailto:Ahmed.Alhammadi@tii.ae}{\{Ahmed.Alhammadi},\href{mailto:carlos-faouzi.bader@tii.ae}{carlos-faouzi.bade\}@tii.ae}).}
\thanks{C. Yuen is with the School of Electrical and Electronics Engineering, Nanyang Technological University, Singapore 639798 (E-mail: \href{mailto:chau.yuen@ntu.edu.sg}{chau.yuen@ntu.edu.sg}).}
\thanks{M. Debbah is with KU 6G Research Center, Khalifa University of Science and Technology, P O Box 127788, Abu Dhabi, UAE
(E-mail: \href{mailto:merouane.debbah@ku.ac.ae}{merouane.debbah@ku.ac.ae}).}

}
\maketitle

\pagestyle{empty}  
\thispagestyle{empty} 

\begin{abstract}
Millimeter-wave (mmWave) multiple-input multiple-output (MIMO) communication with the advanced beamforming technologies is a key enabler to meet the growing demands of future mobile communication.
However, the dynamic nature of cellular channels in large-scale urban mmWave MIMO communication scenarios brings substantial challenges, particularly in terms of complexity and robustness.
To address these issues, we propose a robust gradient-based liquid neural network (GLNN) framework that utilizes ordinary differential equation-based liquid neurons to solve the beamforming problem.
Specifically, our proposed GLNN framework takes gradients of the optimization objective function as inputs to extract the high-order channel feature information, and then introduces a residual connection to mitigate the training burden.
Furthermore, we use the manifold learning technique to compress the search space of the beamforming problem.
These designs enable the GLNN to effectively maintain low complexity while ensuring strong robustness to noisy and highly dynamic channels.
Extensive simulation results demonstrate that the GLNN can achieve 4.15\% higher spectral efficiency than that of typical iterative algorithms, and reduce the time consumption to only 1.61\% that of conventional methods.
\end{abstract}

\section{Introduction}\label{sec:intro}
The sixth-generation (6G) is expected to incorporate a wide-range of communication capabilities that support an enriched and immersive experience, ensure ubiquitous coverage, and enable the new forms of collaboration \cite{gao}. 
Millimeter-wave (mmWave) massive multiple-input multiple-output (MIMO) with the advanced beamforming technologies is recognized as a pivotal solution for 6G to enhance the system throughput \cite{huang6g}. 
However, new challenges come accompanying with the benefits of the mmWave MIMO technology, as mmWave channels are with high propagation loss and typically dominated by the line of sight (LoS) path.
\textcolor{black}{These factors result in a highly dynamic nature of mmWave channels in urban scenarios with a dense presence of small-scale obstacles such as vehicles\cite{6G2}}, which makes it very hard to obtain the channel state information (CSI) accurately \cite{csi2}.
\textcolor{black}{
Therefore, designing an effective
beamforming algorithm in large-scale urban mmWave MIMO
communication scenarios has a substantial challenge.}

To tackle with this challenge, several iterative methods have been introduced \cite{shen2013worst,palhares2021robust}.
\cite{shen2013worst} pioneered the worst-case optimization, while \cite{palhares2021robust} developed this approach with iterative precoders to improve robustness and reduce interference under conditions of imperfect CSI.
However, the complexity of these methods increases greatly with more transmit antennas, challenging their applications in massive MIMO. 
\textcolor{black}{Additionally, the impact mobility on the users has not been fully investigated.}

On the other hand, some researchers have turned to the neural network (NN) based methods for their automatic feature extraction capabilities \cite{huang2020ddpg,gmml,LAGD,zhu2023robust,wang2023energyefficient,wgan,Hasani2022lnn,cfc}.
Specifically, \cite{huang2020ddpg} adopted deep reinforcement learning approaches that employ deep deterministic policy gradients to accommodate continuous action and state spaces. However, such approaches often incur high overhead. 
Recent studies \cite{gmml,LAGD,zhu2023robust,wang2023energyefficient,wgan} have explored deep NN models to reduce the overhead. However, their inherent structural limitations prevent them from capturing time domain information, which limits their effectiveness in processing highly dynamic mmWave channels. \textcolor{black}{To address this challenge, an NN structure named liquid neural network (LNN) was introduced \cite{Hasani2022lnn,cfc}.}
\textcolor{black}{Drawing inspiration from the nervous system of Caenorhabditis elegans, the LNN mimics the time-varying behavior of neural synapses using ordinary differential equations (ODEs), leveraging biologically optimized structures to process noisy 
and highly dynamic data with high efficiency.}

In this paper, we propose the gradient-based liquid neural network (GLNN) method to solve the beamforming problem in mmWave massive MIMO communication. This scheme combines manifold learning techniques and the LNN with a gradient-based learning approach to form a unified framework.
Specifically, the ODE-based structure and learnable parameters enable GLNN to effectively learn from noisy and dynamic CSI. The GLNN employs gradient-based learning to extract the high-order information from the inputs and manifold learning to reduce the optimization space. 
\textcolor{black}{
Differing from previous work \cite{gmml}, the proposed ODE-based GLNN makes several key improvements.
Firstly, unlike the earlier model which uses linear neural layers, the GLNN integrates synapses to form feedback loops of hidden states. These loops utilize historical paths to guide the optimization direction, leveraging the time-domain capabilities of the LNN.
Besides, we replace the complex meta-learning architecture with a simpler single learner model, reducing computational overhead. Finally, we streamline the optimization by using straightforward NN propagation instead of using redundant gradient accumulation.
These novel designs allow the GLNN to achieve the higher performance and stronger robustness to channel estimation errors
(CEE) with a compact NN.}
Unlike traditional NN-based methods, the GLNN only requires a short warm-up period instead of extensive pre-trainings as well. Simulation results demonstrate that the GLNN can achieve 4.15\% higher spectral efficiency (SE) than that of typical iterative algorithms with as low as only 1.61\% of the time consumption.
To support further work, we have made our work open source at \cite{sourcecode}.

\textit{Notation}: Scalars, vectors, and matrices are denoted by $a,\mathbf{a},\mathbf{A}$, respectively. We denote transpose, Hermitian, inverse, $L^2$ norm, Hadamard product, modulus, computational complexity, expectation determinant, trace, all-zero and all-one vectors and identity matrices as $\mathbf{A}^{\mathrm{T}},\mathbf{A}^{\mathrm{H}},\mathbf{A}^{-1},\|\mathbf{A}\|_2,\odot,|\cdot|,\mathcal{O}, \mathbb{E}(\cdot), \det(\mathbf{\cdot}), \mathrm{Tr}(\mathbf{\cdot}),\mathbf{0, 1, I}$, respectively. We denote the complex Gaussian distribution as $\mathcal{CN}(\cdot,\cdot)$.

\section{System Model and Problem Formulation}\label{sec:sys}
In this section, we present the system model, and formulate the object maximization problem in mmWave massive MIMO communication.

Consider a downlink mmWave massive multi-user MIMO (MU-MIMO) communication system where a base station (BS) with $M$ transmit antennas simultaneously serves $K$ users, each of which are equipped with $N_k$ receive antennas. The transmitted signal $\mathbf{u} \in \mathbb{C}^{M\times 1}$ can be represented as
$\mathbf{u}\triangleq \mathbf{w}_ks_k$,
where $\mathbf{w}_k \in \mathbb{C}^{M\times 1}$ denotes the precoding matrix for user $k$, and $s_k\in \mathbb{C}$ assumed as $s_k \sim \mathcal{CN}(0,1)$ denotes the symbol intended for user $k$. We denote the channel matrix from the BS to the user $k$ as $\mathbf{H}_k \in \mathbb{C}^{N_k\times M}$. The additive Gaussian noise vector as $\mathbf{n}_k\in \mathbb{C}^{N_k \times 1}$ with distribution $\mathcal{CN}(\mathbf{0},\sigma^2\mathbf{I})$, where $\sigma^2$ is the variance of the noise.
Therefore, the received signal $\mathbf{y}_k$ at user $k$ can be expressed as
\begin{equation}\label{one_received}
    \mathbf{y}_k =\mathbf{H}_k\mathbf{w}_ks_k+\sum_{j=1,j\neq k}^{K}\mathbf{H}_k\mathbf{w}_j s_j+\mathbf{n}_k,
\end{equation}
where the $s_k$ and $\mathbf{n}_k$ are assumed to be independent for different $k$.
For clarity, we define $N\triangleq \sum_{k=1}^{K}N_k$ as the total number of receive antennas, and denote $\mathbf{y}\triangleq [\mathbf{y}_1^\mathrm{T}, \mathbf{y}_2^\mathrm{T},\cdots,\mathbf{y}_K^\mathrm{T}]^\mathrm{T} \in \mathbb{C}^{N\times 1}$, $\mathbf{W}\triangleq [\mathbf{w}_1, \mathbf{w}_2,\cdots,\mathbf{w}_K] \in \mathbb{C}^{M\times K}$, $\mathbf{s}\triangleq [s_1, s_2,\cdots,s_K]^\mathrm{T} \in \mathbb{C}^{N\times 1}$, and $\mathbf{n}\triangleq [\mathbf{n}_1^\mathrm{T}, \mathbf{n}_2^\mathrm{T},\cdots,\mathbf{n}_K^\mathrm{T}]^\mathrm{T} \in \mathbb{C}^{N\times 1}$.
We make a mild assumption that $\mathbf{H}\triangleq [\mathbf{H}_1^\mathrm{T}, \mathbf{H}_2^\mathrm{T},\cdots,\mathbf{H}_K^\mathrm{T}]^\mathrm{T} \in \mathbb{C}^{N\times M}$ has a full row rank. Therefore, \eqref{one_received} can be rewritten as $
\mathbf{y} = \mathbf{H}\mathbf{W}\mathbf{s}+\mathbf{n}.$

\textcolor{black}{In order to evaluate the robustness of algorithms with inaccurate channel information, we denote the estimated channel as $\mathbf{\hat H}$. We measure the accuracy of channel estimation with CEE in decibels (dB), which is defined as
$\mathrm{CEE} \triangleq 10\log_{10}\left(\frac{\mathbb{E}[\| \mathbf{H} -  \hat{\mathbf{H}} \|_2^2]}{\mathbb{E}[\| \mathbf{H} \|_2^2]}\right)$.}
This indicates that a lower CEE represents a more accurate channel estimation.
\par
\textcolor{black}{
The SE of the system is given by
\begin{equation}\label{SE}
    R=\sum_{k=1}^{K}\alpha_k R_k,
\end{equation}
where the weight $\alpha_k$ represents the priority of the user k. $R_k$ denotes the achievable rate of the user $k$, which is given by
\begin{equation}\label{rkinv}
\begin{split}
        R_k\triangleq \log_2 &\det \bigg(\mathbf{I}+\mathbf{H}_k\mathbf{w}_k(\mathbf{H}_k\mathbf{w}_k)^{\mathrm{H}}\\
        &\times \left(\sum_{j\neq k}^{K} \mathbf{H}_k\mathbf{w}_j(\mathbf{H}_k\mathbf{w}_j)^{\mathrm{H}}+\sigma^2\mathbf{I}\right)^{-1}  \bigg).
\end{split}
\end{equation}
}
\par
The basic problem is to find an optimal $\mathbf{W}$ with a constraint power that maximizes the SE of the system.
In this paper, we use the sum power constraint (SPC) to limit the power of the transmit antennas, which is given as $\mathrm{Tr}(\mathbf{WW}^{\mathrm{H}})\leq P$, where $P$ is the sum power budget of the transmit antennas. Under SPC, the problem can be formulated as
\begin{equation}\label{problem}
    \begin{split}
        \max_{\mathbf{W}}& \ \ R=\sum_{k=1}^{K}\alpha_k R_k\\
        \mathrm{s.t.}&\ \ \mathrm{Tr}(\mathbf{WW}^{\mathrm{H}})\leq P.
    \end{split}
\end{equation}

The function of SE is highly non-convex and non-linear, which makes the optimization problem \eqref{problem} NP-hard. 

\section{Gradient-based Liquid Neural Network}\label{sec:model}
In this section, we present the GLNN scheme which combines manifold learning techniques and LNN with a gradient-based learning approach to form a cohesive framework.

\subsection{Manifold Learning Technique}
For massive MU-MIMO communication system, the BS is usually equipped with a large number of transmit antennas. \textcolor{black}{This makes the search space of \eqref{problem} extremely large, resulting in demanding and expensive training progress.} Additionally, the number of transmit antennas at the BS is typically much larger than the number of receive antennas at the users \cite{BjMIMO}. Therefore, an efficient algorithm may have a search space that is independent of $M$ \cite{rwmmse}.
Considering these factors, we utilize the manifold learning technique \cite{gmml} to compress the search space for \eqref{problem}.
With the help of the manifold learning technique, the optimal $\mathbf{W}$ can be compressed as
$\mathbf{W}=\mathbf{H}^{\mathrm{H}}\mathbf{X}$,
where $\mathbf{X}\in \mathbb{C}^{N\times K}$ is the base matrix for the optimal $\mathbf{W}$. Given $M\gg N$, this greatly reduces the search space from $\mathbb{C}^{M\times K}$ to a manifold of $\mathbb{C}^{N\times K}$. The proposed GLNN exploits this feature by optimizing the low-dimensional base matrix instead of optimizing the precoding matrix directly.

\subsection{LNN Architecture}
LNNs are networks that are constructed by linear first-order dynamical systems and controlled through nonlinear interconnected gates \cite{Hasani2022lnn}.
These networks are based on ODEs without the necessity for a solver and can model dynamical systems whose time constants are liquid in relation to their hidden states. A basic ODE can be expressed as
\begin{equation}\label{basicde}
   \frac{\mathrm{d} p(t)}{\mathrm{d} t}=-\frac{p(t)}{\tau}+S(t),
\end{equation}
where $S(t)=f(i(t))(A-x(t))$ represents the synaptic current between neurons, and $p(t),A,\tau$ represents the hidden state, bias and time constant, respectively. To derive a closed-form solution, the hidden states of a layer of ODE-based liquid neurons can be determined by an initial value problem, which can be expressed as
\begin{equation}\label{ivp}
\begin{split}
        \frac{\mathrm{d}\mathbf{p}(t)}{\mathrm{d}t}=&-[\mathbf{o}_\tau+\mathbf{f}(\mathbf{p}(t),\mathbf{i}(t),\theta)\odot \mathbf{p}(t)]\\
        &+\mathbf{a}\odot \mathbf{f}(\mathbf{p}(t),\mathbf{i}(t),\theta),
\end{split}
\end{equation}
where $\mathbf{p}(t)\in \mathbb{R}^{D\times 1},\mathbf{i}(t)\in \mathbb{R}^{C\times 1},\mathbf{o}_\tau\in \mathbb{R}^{D\times 1},\mathbf{a}\in \mathbb{R}^{D\times 1},\mathbf{f}$ are the hidden states of a layer of $D$ neurons (i.e. NN head), external inputs with $C$ features, time constant parameter, bias and the function of the NN head with parameters $\theta_{\mathbf{f}}$,
respectively. Given a continuous system described in \eqref{ivp}, the closed-form solution can be approximated as
\begin{equation}\label{ltc}
\begin{split}
        \mathbf{p}(t)&=(\mathbf{p}(0)-\mathbf{a})\odot e^{-\mathbf{o}_\tau t-\int_0^t\mathbf{f}(\mathbf{i}(s),\theta_{\mathbf{f}})\mathrm{d}s}+\mathbf{a}\\
         &\approx \mathbf{b}\odot e^{-[\mathbf{o}_\tau +\mathbf{f}(\mathbf{i}(t),\theta_{\mathbf{f}})]t}\odot \mathbf{f}(-\mathbf{i}(t),\theta_{\mathbf{f}})+\mathbf{a},
\end{split}
\end{equation}
where $(\mathbf{p}(0)-\mathbf{a})$ is denoted as $\mathbf{b}\in \mathbb{R}^{D\times 1}$ to allow explicit derivation of $\mathbf{p}(t)$.
\textcolor{black}{\eqref{ltc} provides an approximation that eliminated the integration, which can significantly reduce the computational complexity.}
However, the exponential term in \eqref{ltc} causes the system to quickly converge to $\mathbf{a}$.
To address this, we replace the exponential decay term with a smooth nonlinearity introduced by sigmoid gates. 
\textcolor{black}{Finally, to enhance the flexibility of the GLNN, we make both $\mathbf{a}$ and $\mathbf{b}$ trainable by replacing them with NN heads $\mathbf{g}$ and $\mathbf{h}$, respectively.}
\textcolor{black}{To better consider the temporal dependencies, the input to each liquid neuron not only includes the current feature input $\mathbf{I}$, but also the previous hidden state $\mathbf{p}$.}
\begin{figure}[t]
	\begin{center}
		\centerline{\includegraphics[width=0.65\linewidth]{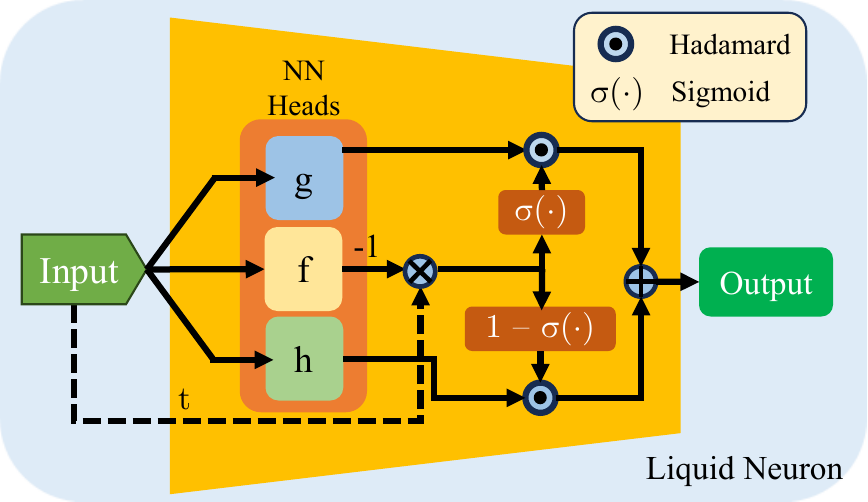}}  \vspace{-0mm}
		\captionsetup{font=footnotesize, name={Fig.}, labelsep=period} 
		\caption{\, Structure of a liquid neuron.}
		\label{fig:cfc}\vspace{-8mm}
	\end{center}
\end{figure}
\begin{figure}[t]
	\begin{center}
		\centerline{\includegraphics[width=0.9\linewidth]{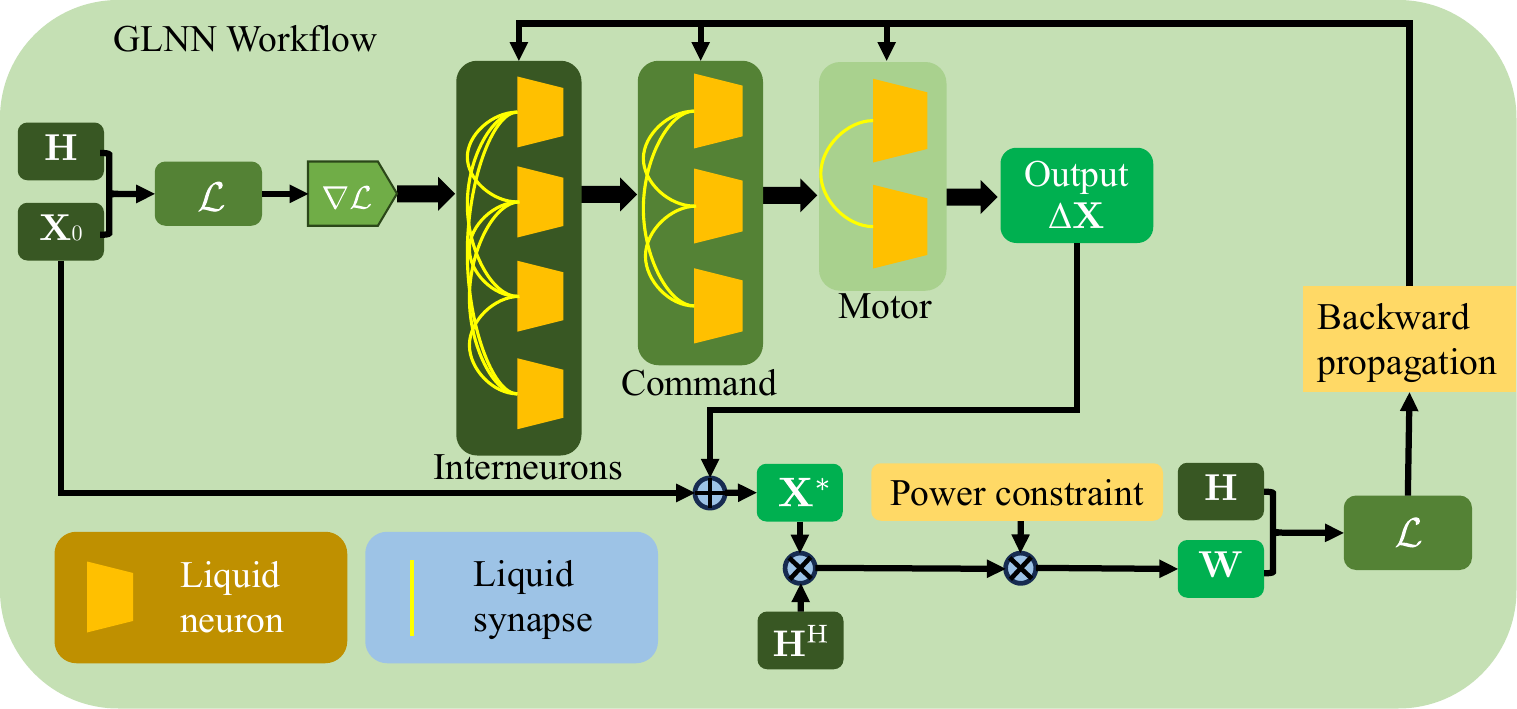}}  \vspace{-0mm}
		\captionsetup{font=footnotesize, name={Fig.}, labelsep=period} 
		\caption{\, \textcolor{black}{Structure of GLNN.}}
		\label{fig:glnn}\vspace{-8mm}
	\end{center}
\end{figure}
The closed-form continuous-time model for a layer of liquid neurons illustrated in Fig. \ref{fig:cfc} can be presented as
\begin{equation}\label{cfc}
\begin{split}
    \mathbf{p}(t)=&\sigma(-\mathbf{f}(\mathbf{p},\mathbf{I};\theta_{\mathbf{f}})t)\odot \mathbf{g}(\mathbf{p},\mathbf{I};\theta_{\mathbf{g}})\\
    &+[\mathbf{1}-\sigma(-\mathbf{f}(\mathbf{p},\mathbf{I};\theta_{\mathbf{f}})t)]\odot \mathbf{h}(\mathbf{p},\mathbf{I};\theta_{\mathbf{h}}),
\end{split}
\end{equation}
where $\mathbf{\theta}_{\mathbf{g}},\mathbf{\theta}_{\mathbf{h}}$ denote the parameters of NN heads with functions of $\mathbf{g},\mathbf{h}$, respectively.

\begin{algorithm}[t]
\caption{GLNN Workflow }
\label{alg:glnn}
\begin{algorithmic}[1]
\Procedure{GLNN}{$\mathbf{H}$}
    \State Randomly Initialize $\mathbf{\theta},\mathbf{X}_0$.
    \State Initialize $\mathbf{W}_0$ by $\mathbf{W}_0=\mathbf{H}^{\mathrm{H}}\mathbf{X}_0$
    \For{$i\leftarrow 1,2,\cdots,N_e$}
        \State Calculate $\mathcal{L}_0$ with $\mathbf{H},\mathbf{W}_0$ by \eqref{loss}
        \State $\mathbf{\Delta X}=\mathrm{LNN}(\nabla \mathcal{L}_0,\theta)$
        \State Calculate $\mathbf{W}$ by \eqref{pwrcst}
        \State Calculate the $\mathcal{L}$ as \eqref{loss}
        \State Calculate and record the $R$ as \eqref{SE}
        \State Update $\theta$ as \eqref{bwdppg}
    \EndFor
    \State \Return $\mathbf{W}$
\EndProcedure
\end{algorithmic}
\end{algorithm}
\subsection{Gradient-based Optimization}
Prior deep learning (DL)-based beamforming schemes typically feed the raw channel matrix $\mathbf{H}$ into the NNs, and use the output directly as the precoding matrix $\mathbf{W}$. However, due to the high non-convexity of the optimization space, traditional NN-based architectures can hardly extract high-order information from the this space \cite{gmml, wang2023energyefficient}, which can reduce the overall performance. 
To tackle with this, we introduce the gradient-based optimization, which is a technique that feeds gradients of the optimization object into the NNs. 
This technique combines the ability of the NNs to extract time domain features of the optimization space with the high-order information extracted from the gradient-as-input mechanism, potentially leading to improved optimization performance.
However, this combination may lead to difficulties in updating the parameters of the NNs.
Inspired by the conception of residual learning technique in \cite{resnet}, we treats the output as the residual $\mathbf{\Delta X}$. This output is then used to update $\mathbf{X}$ by $\mathbf{X}^*=\mathbf{X}+\mathbf{\Delta X}$, where $\mathbf{X}^*$ denotes the updated base matrix.

\subsection{GLNN Framework}
This section introduces the GLNN framework as described in Algorithm \ref{alg:glnn} and Fig. \ref{fig:glnn}
\subsubsection{Forward Propagation}
This procedure updates the precoding matrix. We first initialize the base matrix as $\mathbf{X}_0$ and calculate the loss function and its gradient with respect to $\mathbf{X}_0$ as $\nabla \mathcal{L}\in \mathbb{C}^{N\times K}$. \textcolor{black}{Then we consider the $\nabla \mathcal{L}$ as a batch of vectors of dimension $K$ and feed them into three layers of liquid neurons that are fully connected by liquid synapses, which are named interneurons, command, and motor, respectively.} Both the input and the output are decomposed into real and imaginary parts, thus both the interneurons layer and the motor layer contains $2K$ liquid neurons. Finally, considering the power constraints in \eqref{problem}, we can compute the precoding matrix as
\begin{equation}\label{pwrcst}
    \mathbf{W}=\sqrt{\frac{P}{\mathrm{Tr}(\mathbf{H}^{\mathrm{H}} \mathbf{X}^* (\mathbf{H}^{\mathrm{H}} \mathbf{X}^*)^{\mathrm{H}})}}\cdot \mathbf{H}^{\mathrm{H}} \mathbf{X}^*,
\end{equation}

\subsubsection{Backward Propagation}
This procedure updates the LNN parameters.
We noticed that NN-based algorithms usually have a tendency to provide signals to only one user, which introduces a fairness issue and limits the overall SE.
Therefore, diverging from traditional methods that directly use $-R$ as the loss function, our strategy includes a penalty term to mitigate rate disparities among users by adding the variance of $R_k$ for all users, denoted as $\text{Var}(R)$.
To counterbalance the potentially excessive penalties from the penalty term that may hinder the optimization of GLNN, we introduce an incentive mechanism within our loss function. This mechanism initially encourages the NN to undertake a straightforward search at the early stage, which prevents the network from overly prioritizing fairness and discouraging optimization efforts. 
The loss function can be denoted as
\begin{equation}\label{loss}
    \mathcal{L}=-R+\beta \cdot \mathrm{Var}(R)+\gamma \cdot \mathrm{ReLU} (\lambda \cdot K-R),
\end{equation}
where $\mathcal{L},\beta,\gamma,\lambda$ denote the loss function, penalty rate, incentive rate and threshold, respectively.
\textcolor{black}{The $\lambda$ in \eqref{loss} serves as a threshold of the incentive mechanism, which value can be determined with linear search in the warm-up period.}
Then the backward propagation is conducted with Adam optimizer, as
\begin{equation}\label{bwdppg}
    \theta^*=\theta+\alpha \cdot \mathrm{Adam(\nabla_\theta \mathcal{L},\theta}),
\end{equation}
where $\theta,\theta^*$ are NN parameters  $\{\mathbf{\theta}_{\mathbf{f}},\mathbf{\theta}_{\mathbf{g}},\mathbf{\theta}_{\mathbf{h}}\}$ before and after update, respectively, and $\alpha$ is the learning rate.

\section{Simulation Results}\label{sec:simulation}
\textcolor{black}{In this section, we discuss the performance of the proposed GLNN method through simulation results. We construct the channel matrix using BS3 from scenario O1 of the open source DeepMIMO dataset \cite{deepmimo} with parameters $M=64,N_k=2,K=4,f_c=28 \mathrm{\ GHz},\alpha_k=1,k=1,2,\cdots,K$, where $f_c$ is the central frequency.}
We establish a warm-up period containing $N_r$ samples for DL-based methods to learn knowledge from continuous samples over a period of time, and set $\alpha=0.01,\beta=0.3,\gamma=0.7,\lambda=2.5,N_e=3,N_r=500,\sigma^2=0 \mathrm{\ dBm},P=10\mathrm{\ dBm}$ as the default hyperparameters. The command layer in the GLNN comprises 30 liquid neurons. We generate 500 samples with continuously changing positions and average the results across all the samples. The inaccuracy of channel estimation $\mathbf{\hat H}-\mathbf{H}$ is assumed to follow a zero-mean white Gaussian distribution.
We run all the simulations on a computer with an Intel i7-12700H CPU using Pytorch 2.1.2 and Python 3.11.
We compare GLNN to several baselines as listed below.

\begin{itemize}
    \item \textbf{Baseline 1} (WMMSE): An iterative scheme from \cite{wmmse}.
    \item \textbf{Baseline 2} (LAGD): A DL-based model from \cite{LAGD}.
    \item \textbf{Baseline 3} (LSTM): A DL-based model employing a straightforward LSTM approach, directly mapping the $\mathbf{\hat{H}}$ to $\mathbf{W}$ without the use of additional techniques.
    \item \textbf{Baseline 4} (Upper Bound): To determine the potential maximal SE, we run the GLNN and all the baselines on perfect channel information sufficient times (i.e. 100 times) with independent random initialization and record the highest one as the upper bound.
\end{itemize}

\subsection{Performance Evaluation}
In this subsection, we evaluate the performance of the GLNN by comparing the SE with the baselines.
\subsubsection{Impact of $P$}
\begin{figure}[t]
	\begin{center}
		\centerline{\includegraphics[width=0.8\linewidth]{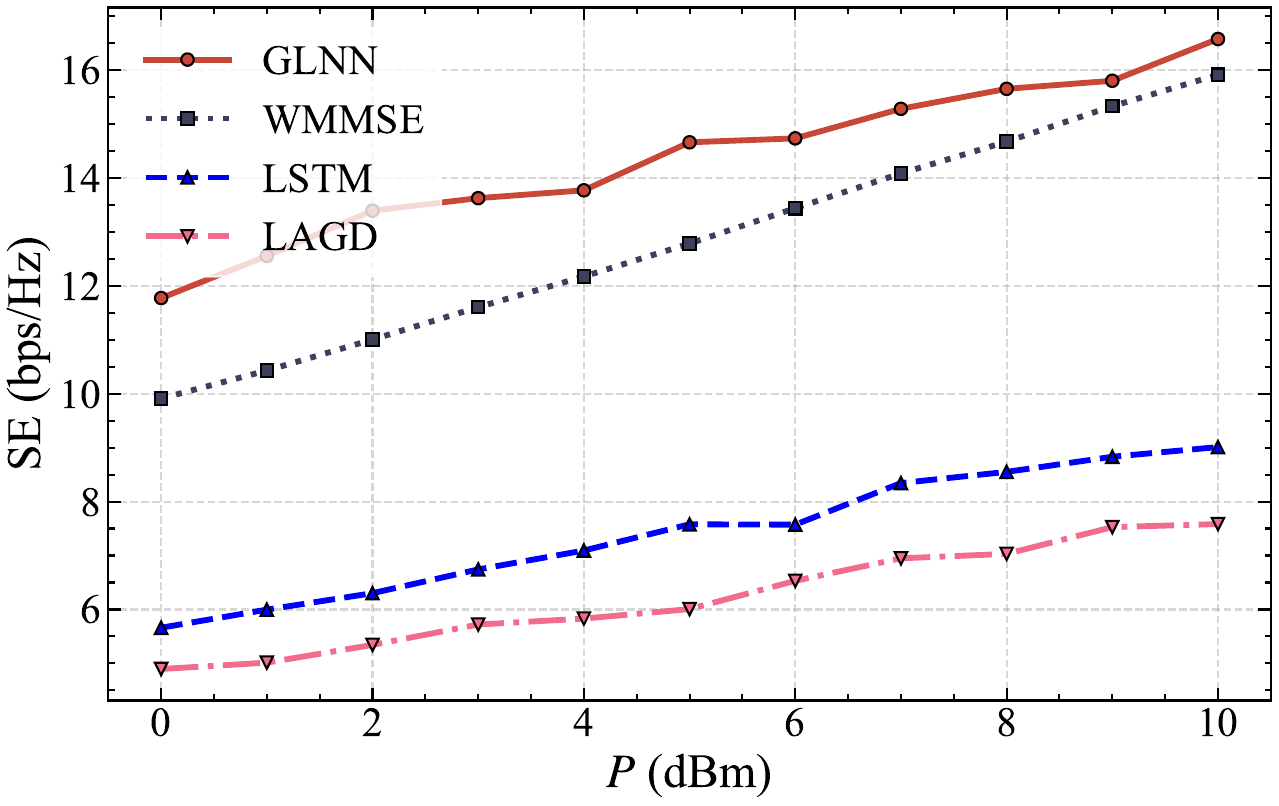}}  \vspace{-0mm}
		\captionsetup{font=footnotesize, name={Fig.}, labelsep=period} 
		\caption{\, SE vs. the transmit power under perfect channel information.}
		\label{fig:snr}\vspace{-6mm}
	\end{center}
\end{figure}

In Fig. \ref{fig:snr}, we present the SE of the proposed GLNN alongside the baselines under perfect channel information, under conditions of fixed $\sigma^2$ and varying $P$.
The results show that the SE of all the algorithms increases with an increasing $P$, and the proposed GLNN outperforms all the baselines, being 4.15\% higher than the WMMSE when $P$ is 10 dBm.
The underperformance of the LSTM and LAGD than the GLNN is due to design limitations. Specifically, the inability of the LSTM to effectively extract high-order information hampers optimization in complex spaces.
Meanwhile, the absence of an ODE-based architecture in the LAGD reduces its adaptability in dynamic scenarios, despite its theoretical capabilities in managing high-dimensional data.

\begin{figure}[t]
	\begin{center}
		\centerline{\includegraphics[width=0.8\linewidth]{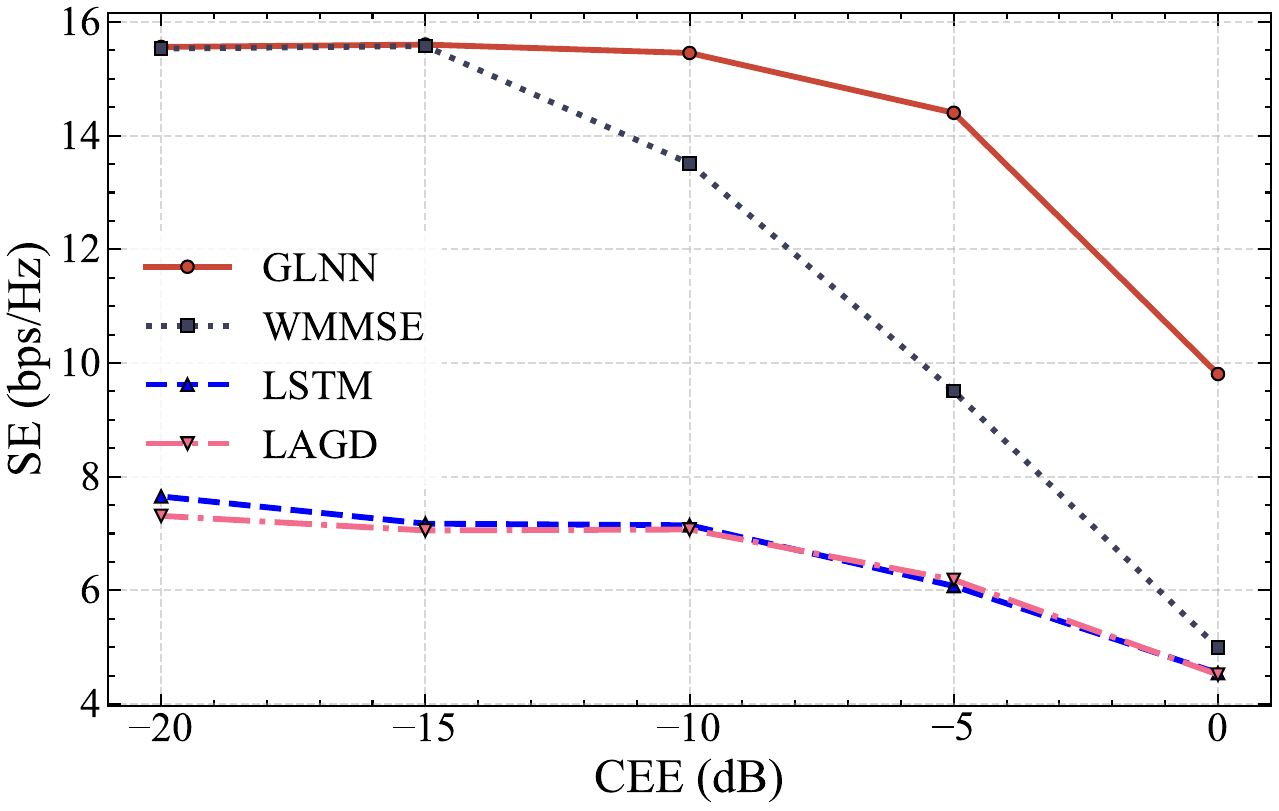}}  \vspace{-0mm}
		\captionsetup{font=footnotesize, name={Fig.}, labelsep=period} 
		\caption{\, SE vs CEE under $P=10$ dBm.}
		\label{fig:cee}\vspace{-9mm}
	\end{center}
\end{figure}

\subsubsection{Impact of CEE}
In Fig. \ref{fig:cee}, we fix the $P$ at 10 dBm and evaluate the performance of the algorithms as the CEE increases from -20 dB to 0 dB. The results show that both the WMMSE and the GLNN exhibit high SE at low CEE levels.  However, while the WMMSE demonstrates reduced robustness against high CEE, the GLNN shows strong robustness, exhibiting a 96.59\% higher SE at a CEE of 0 dB.
This advantage is primarily due to its fully connected neural synapses and its ODE-based design, which allows GLNN to extract features from noisy inputs more effectively.
\subsubsection{Dynamic Scenarios}
\begin{figure}[t]
	\begin{center}
		\centerline{\includegraphics[width=0.82\linewidth]{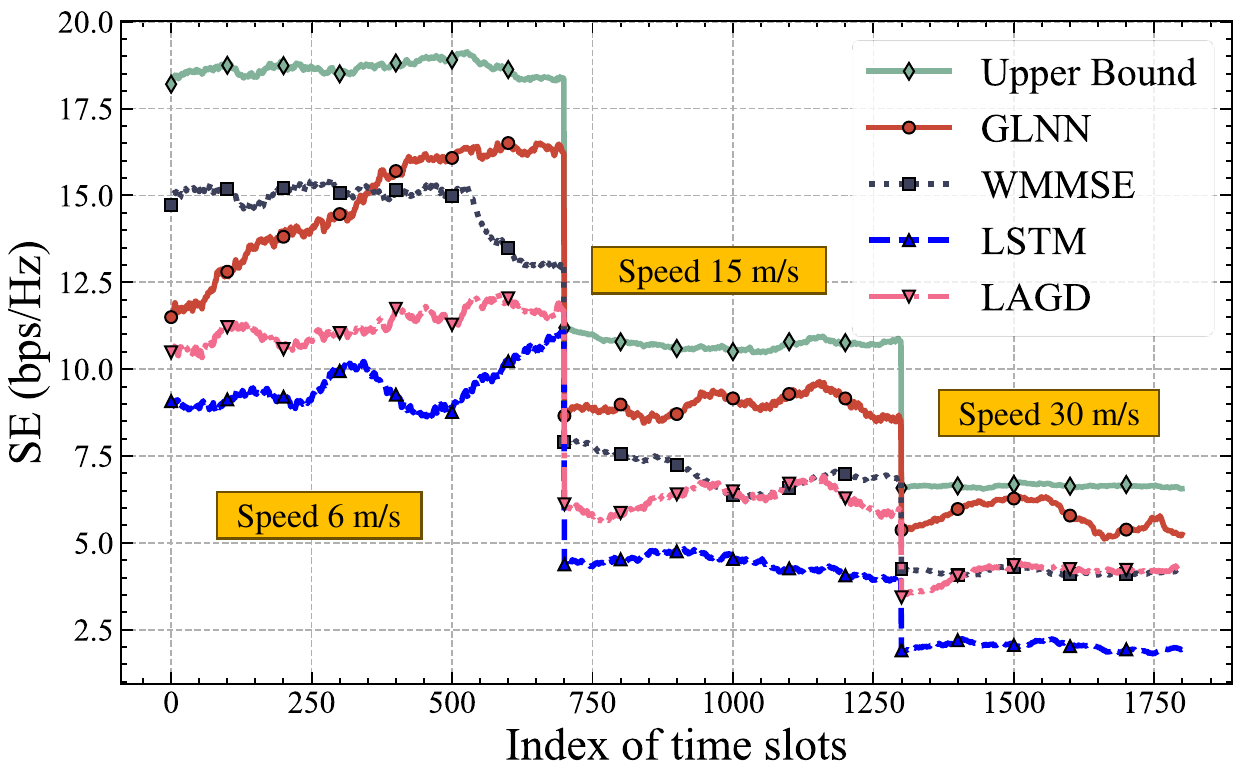}}  \vspace{-2mm}
		\captionsetup{font=footnotesize, name={Fig.}, labelsep=period} 
		\caption{\, Dynamic simulation results with CEE = -10 dB and without warm-up.}
		\label{fig:dynresult}\vspace{-6mm}
	\end{center}
\end{figure}

\textcolor{black}{In Fig. \ref{fig:dynresult}, we remove the warm-up phase for the LAGD, the LSTM and the GLNN and run all algorithms under a -10 dB CEE to more closely simulate real-world online performance.} We provide three phases of speed, including 6 m/s, 15 m/s, and 30 m/s, with each phase comprising 700, 600, and 500 time slots, respectively. We smoothed the results in each phase for clarity. 
The GLNN model rapidly surpasses WMMSE after a short period and subsequently maintains higher SE over all baselines.
Notably, the gap between the Upper Bound and the DL-based models widens as changing to phases of higher speeds. However, this increased gap swiftly closes as the GLNN adapts after just a few iterations, demonstrating its high adaptability and performance in dynamically changing environments.
This is due to the multi-head design of liquid neurons in the GLNN, which replaces the bias and zero-state with trainable parameters and thus enables GLNN to adapt quickly to scenario changes.
\begin{figure}[t]
	\begin{center}
		\centerline{\includegraphics[width=0.82\linewidth]{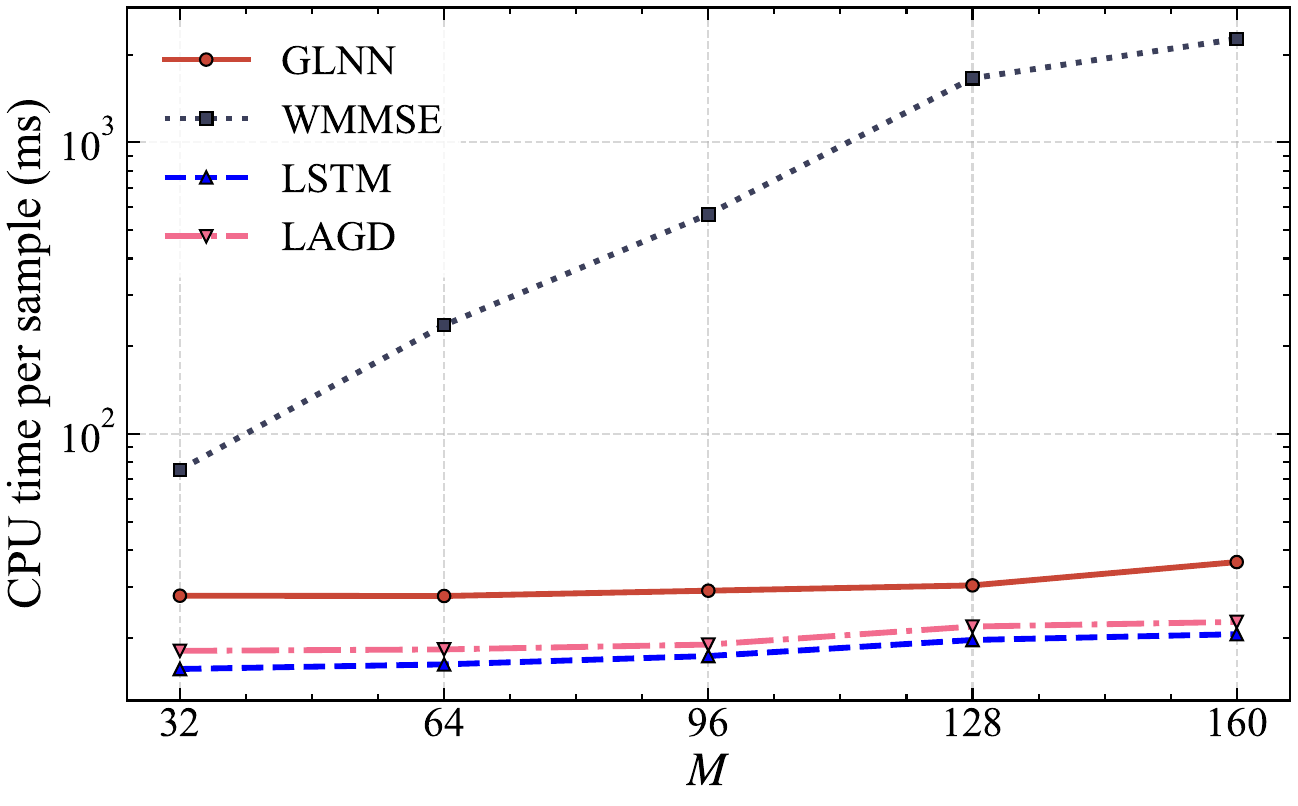}}  \vspace{-2mm}
		\captionsetup{font=footnotesize, name={Fig.}, labelsep=period} 
		\caption{\, Time for optimization with different $M$.}
		\label{fig:time}\vspace{-8mm}
	\end{center}
\end{figure}

\subsection{Time Consumption Evaluation}
In this subsection, we compare the computational complexity of GLNN and the baseline algorithms as the number of BS transmit antennas increases. According to Fig. \ref{fig:time}, the time cost of WMMSE rises rapidly with more antennas, while the rise of GLNN is slower, reaching a ratio of 1.61\% compared to the WMMSE when $M=160$. On the other hand, despite their lower complexity, LAGD and LSTM offer reduced SE, which limits their application.
\textcolor{black}{
The main source of complexity in the GLNN are the LNN and the computation of $R$.
The LNN features three layers of liquid neurons, each introducing complexity linear to $K$, with a batch size of $N$.
Thus, the complexity of the LNN is $\mathcal{O}(NK)$ [15].
The calculation of $R$ involves $K$ instances of $\mathbf{H}_k\mathbf{w}_j(\mathbf{H}_k\mathbf{w}_j)^{\mathrm{H}}$ and $K$ determinant calculation for $N_k\times N_k$ matrices.
Therefore, the complexity to calculate $R$ is $\mathcal{O}(MNK+M^2K+N_k^3)$.
Since these operations are repeated $N_e$ times per sample, the total complexity of GLNN is $\mathcal{O}(N_e(MNK+M^2K+N_k^3))$.
}
Table \ref{tab:complexity}  provides a comparison of the computational complexity between GLNN and the baselines, where $I_\mu,I_{\mathbf{W}}$ are the iteration numbers of the bisection searches and the three-step update loops in the WMMSE, respectively.

\section{Conclusion}\label{sec:conclusion}
In this paper, a robust beamforming method named GLNN was presented. Specifically, the GLNN integrated the manifold learning technique and the ODE-based LNN with the gradient-based learning framework. Simulation results showed that the GLNN outperforms the baselines by 4.15\%, and demonstrated the strong robustness against channel estimation inaccuracies while reducing the time consumption to only 1.61\% that of conventional methods.
\begin{table}[t]
\centering
\caption{Comparison of Computational Complexity}\vspace{-1mm}
\begin{tabular}{c c} 
\toprule
Algorithm & Time Complexity \\ [0.2ex] 
\midrule
GLNN & $\mathcal{O}(N_e(MNK+M^2K+N_k^3))$\\
WMMSE & $\mathcal{O}(I_\mu I_{\mathbf{W}}(MN^2++MNK+M^3K^2+NN_k^2))$\\
LAGD & $\mathcal{O}(N_e(M^2N+N_k^3))$\\
LSTM & $\mathcal{O}(N_e(MNK+M^2K+N_k^3))$\\
\bottomrule
\end{tabular}\vspace{-3mm}
\label{tab:complexity}
\end{table}

\bibliographystyle{IEEEtran}
\bibliography{lnn}

\begin{thebibliography}{10}
\providecommand{\url}[1]{#1}
\csname url@samestyle\endcsname
\providecommand{\newblock}{\relax}
\providecommand{\bibinfo}[2]{#2}
\providecommand{\BIBentrySTDinterwordspacing}{\spaceskip=0pt\relax}
\providecommand{\BIBentryALTinterwordstretchfactor}{4}
\providecommand{\BIBentryALTinterwordspacing}{\spaceskip=\fontdimen2\font plus
\BIBentryALTinterwordstretchfactor\fontdimen3\font minus
  \fontdimen4\font\relax}
\providecommand{\BIBforeignlanguage}[2]{{%
\expandafter\ifx\csname l@#1\endcsname\relax
\typeout{** WARNING: IEEEtran.bst: No hyphenation pattern has been}%
\typeout{** loaded for the language `#1'. Using the pattern for}%
\typeout{** the default language instead.}%
\else
\language=\csname l@#1\endcsname
\fi
#2}}
\providecommand{\BIBdecl}{\relax}
\BIBdecl

\bibitem{gao}
Y.~Wang \emph{et~al.}, ``Transformer-empowered {6G} intelligent networks: From
  massive {MIMO} processing to semantic communication,'' \emph{IEEE Wirel.
  Commun.}, vol.~30, no.~6, pp. 127--135, 2023.

\bibitem{huang6g}
C.~Huang \emph{et~al.}, ``Holographic {MIMO} surfaces for {6G} wireless
  networks: Opportunities, challenges, and trends,'' \emph{IEEE Wirel.
  Commun.}, vol.~27, no.~5, pp. 118--125, 2020.

\bibitem{6G2}
W.~Jiang \emph{et~al.}, ``The road towards {6G}: A comprehensive survey,''
  \emph{IEEE Open J. Commun. Soc.}, vol.~2, pp. 334--366, 2021.

\bibitem{csi2}
W.~Xu \emph{et~al.}, ``Robust beamforming with partial channel state
  information for energy efficient networks,'' \emph{IEEE J. Sel. Areas
  Commun.}, vol.~33, no.~12, pp. 2920--2935, 2015.

\bibitem{shen2013worst}
H.~Shen \emph{et~al.}, ``A worst-case robust {MMSE} transceiver design for
  nonregenerative mimo relaying,'' \emph{IEEE Trans. Wirel. Commun.}, vol.~13,
  no.~2, pp. 695--709, 2013.

\bibitem{palhares2021robust}
V.~M. Palhares, A.~R. Flores, and R.~C. De~Lamare, ``Robust {MMSE} precoding
  and power allocation for cell-free massive mimo systems,'' \emph{IEEE Trans.
  Veh. Tech.}, vol.~70, no.~5, pp. 5115--5120, 2021.

\bibitem{huang2020ddpg}
C.~Huang \emph{et~al.}, ``Multi-hop {RIS}-empowered terahertz communications: A
  {DRL}-based hybrid beamforming design,'' \emph{IEEE J. Sel. Areas Commun.},
  vol.~39, no.~6, pp. 1663--1677, 2021.

\bibitem{gmml}
F.~Zhu \emph{et~al.}, ``{Robust Beamforming for RIS-aided Communications:
  Gradient-based Manifold Meta Learning},'' \emph{IEEE Trans. Wirel. Commun.},
  2024.

\bibitem{LAGD}
Z.~Yang \emph{et~al.}, ``A learning-aided flexible gradient descent approach to
  {MISO} beamforming,'' \emph{IEEE Wirel. Commun. Lett.}, vol.~11, no.~9, pp.
  1895--1899, 2022.

\bibitem{zhu2023robust}
F.~Zhu \emph{et~al.}, ``Robust millimeter beamforming via self-supervised
  hybrid deep learning,'' in \emph{2023 31st Eur. Signal Process. Conf.
  (EUSIPCO)}.\hskip 1em plus 0.5em minus 0.4em\relax IEEE, 2023.

\bibitem{wang2023energyefficient}
X.~Wang \emph{et~al.}, ``Energy-efficient beamforming for {RISs-aided}
  communications: Gradient based meta learning,'' in \emph{Proc. 2024 IEEE Int.
  Conf. Commun. (ICC)}, Jun. 2024.

\bibitem{wgan}
F.~Zhu \emph{et~al.}, ``Beamforming inferring by conditional {WGAN-GP} for
  holographic antenna arrays,'' \emph{IEEE Wirel. Commun. Lett.}, vol.~13,
  no.~7, pp. 2023--2027, 2024.

\bibitem{Hasani2022lnn}
R.~Hasani \emph{et~al.}, ``Closed-form continuous-time neural networks,''
  \emph{Nat. Mach. Intell.}, vol.~4, no.~11, pp. 992--1003, 2022.

\bibitem{cfc}
M.~Chahine \emph{et~al.}, ``Robust flight navigation out of distribution with
  liquid neural networks,'' \emph{Sci. Robot.}, vol.~8, no.~77, p. eadc8892,
  2023.

\bibitem{sourcecode}
X.~Wang, ``{GLNN},'' \textit{\url{https://github.com/tp1000d/GLNN}}, 2024.

\bibitem{BjMIMO}
E.~Björnson \emph{et~al.}, ``Massive {MIMO} is a reality—what is next?: Five
  promising research directions for antenna arrays,'' \emph{Digit. Signal
  Process.}, vol.~94, pp. 3--20, 2019.

\bibitem{rwmmse}
X.~Zhao \emph{et~al.}, ``Rethinking {WMMSE}: Can its complexity scale linearly
  with the number of {BS} antennas?'' \emph{IEEE Trans. Signal Process.},
  vol.~71, pp. 433--446, 2023.

\bibitem{resnet}
K.~He \emph{et~al.}, ``Deep residual learning for image recognition,'' in
  \emph{Proc. IEEE Conf. Comput. Vis. Pattern Recognit. (CVPR)}, 2016.

\bibitem{deepmimo}
A.~Alkhateeb, ``Deepmimo: A generic deep learning dataset for millimeter wave
  and massive {MIMO} applications,'' in \emph{Proc. of Information Theory and
  Applications Workshop (ITA)}, San Diego, CA, Feb 2019, pp. 1--8.

\bibitem{wmmse}
Q.~Shi \emph{et~al.}, ``An iteratively weighted {MMSE} approach to distributed
  sum-utility maximization for a {MIMO} interfering broadcast channel,''
  \emph{IEEE Trans. Signal Process.}, vol.~59, no.~9, pp. 4331--4340, Sept.
  2011.

\end{thebibliography}
\vspace{12pt}
\end{document}